\documentstyle[preprint,aps,prd,epsfig]{revtex}

\begin{document}
\draft

\title{\bf Quantum Fluctuations of Radiation Pressure }

\author{Chun-Hsien Wu \footnote{e-mail: wu@cosmos.phy.tufts.edu} 
        and L. H. Ford \footnote{e-mail: ford@cosmos.phy.tufts.edu}}

\address{Institute of Cosmology, Department of Physics and Astronomy\\
	Tufts University, Medford, MA 02155, USA}

\date{\today}
\maketitle
\tightenlines

\begin{abstract}
Quantum fluctuations of electromagnetic radiation pressure are 
discussed. We
use an approach based on the quantum stress tensor to calculate the 
fluctuations
in velocity and position of a mirror subjected to electromagnetic 
radiation.
Our approach reveals that radiation pressure fluctuations are due to 
a cross
term between vacuum and state dependent terms in a stress tensor 
operator product.
Thus observation of these fluctuations would entail experimental 
confirmation
of this cross term. We first analyze the pressure fluctuations on
a single, perfectly reflecting mirror, and then study the case of
an interferometer. This involves a study of the effects of multiple bounces
in one arm, as well as the correlations of the pressure fluctuations
between arms of the interferometer.
 In all cases, our results are consistent with
those previously obtained by Caves using different mehods.
\end{abstract}

\section{Introduction \label{sec: Intro}}

Classically, a beam of light falling on a mirror exerts a force and the 
force can be written as the integral of the Maxwell stress tensor. 
When we treat this problem quantum mechanically, then the force 
undergoes fluctuations. This is a necessary consequence of the fact
that physically realizable quantum states are not eigenstates of the
stress tensor operator. These radiation pressure fluctuations play an
important role in limiting the sensitivity of laser interferometer
detectors of gravitational radiation, as was first analyzed by 
Caves~\cite{Caves1,Caves2}. His approach is based on the photon number 
fluctuations in a coherent state, and we will refer to it as the
photon number approach. The purpose of this paper is to examine radiation
pressure fluctuations using the quantum stress tensor. This requires
the correlation function of a pair of stress tensor operators, which
will be discussed in Sect.~\ref{sec: EMTF}. There it is shown that
the correlation function can be decomposed into three parts, a term which is 
fully normal-ordered,  a state independent vacuum term, and
a ``cross term'' which can be viewed as an interference term between
the vacuum fluctuations and the matter content of the quantum state.
 It is this cross term
which will be of greatest interest in this paper, as it is responsible
for the radiation pressure fluctuations in a coherent state. These
fluctuations will be discussed in Sect.~\ref{sec: QFRP} for the case
of a laser beam impinging upon a single, perfectly reflecting mirror.
The analysis will be done first using the photon number approach, and then
using the stress tensor approach. In the latter case, we show how the 
calculations may be performed in coordinate space, where an integration over
space and time is needed to remove a singularity in the cross term.
We then show how to obtain the same result more simply using a othonormal
basis of wavepacket modes. In Sect.~\ref{sec: NES}, we examine the case
of a single mode number eigenstate, and show that the radiation pressure
fluctuations vanish. In the stress tensor approach, this arises from a
cancellation between the cross term and the fully normal ordered term.
In Sect.~\ref{sec: Noise}, we turn to the discussion of an interferometer.
We first study the effects of mutiple bounces in a single arm, and reproduce
the result~\cite{Caves1,Caves2} that the effect of the radiation 
pressure fluctuations grows
as the square of the number of bounces. We also show how our approach may 
be used to discuss the situation where the interferometer arms are 
Fabry-Perot cavities. Finally, we discuss the correlation between the
fluctuations in the two interferometer arms, and show why they are in 
fact uncorrelated. Our results are summarized and discussed in 
Sect.~\ref{sec:final}.

\section{Energy-momentum tensor fluctuations }
\label{sec: EMTF}

It is well known that stress tensor operators can be renormalized by 
normal ordering: 
\begin{equation}
:T_{\mu \nu }:=T_{\mu \nu }-\langle T_{\mu \nu }\rangle _{0}\, ,
\end{equation}
 which is subtraction of the Minkowski vacuum expectation value. 
However, the
quantity \( \langle :T_{\mu \nu }(x)::T_{\rho \sigma }(x'):\rangle  \) 
is still
divergent in the limit that \( x' \rightarrow x \). The divergent part 
of this
quantity can be decomposed into a state-independent part and a 
state-dependent
part. To do so, we may use the following identity, which follows from 
Wick's
theorem 
\begin{eqnarray}
:\phi _{1}\phi _{2}::\phi _{3}\phi _{4}: & = & :\phi _{1}\phi 
_{2}\phi _{3}\phi _{4}:\nonumber \\
 &  & +:\phi _{1}\phi _{3}:\langle \phi _{2}\phi _{4}\rangle 
_{0}+:\phi _{1}\phi _{4}:\langle \phi _{2}\phi _{3}\rangle 
_{0}\nonumber \\
 &  & +:\phi _{2}\phi _{3}:\langle \phi _{1}\phi _{4}\rangle 
_{0}+:\phi _{2}\phi _{4}:\langle \phi _{1}\phi _{3}\rangle 
_{0}\nonumber \\
 &  & +\langle \phi _{1}\phi _{3}\rangle _{0}\langle \phi _{2}\phi 
_{4}\rangle _{0}+\langle \phi _{1}\phi _{4}\rangle _{0}\langle \phi 
_{2}\phi _{3}\rangle _{0}\, .\label{eq:phiprod} 
\end{eqnarray}
Here the \( \phi _{i} \) are free bosonic fields and \( \langle \, 
\rangle _{0} \)
denotes the Minkowski vacuum expectation value. The first term is 
fully normal-ordered,
the next four are cross terms and the final two are pure vacuum 
terms. The physics
of these various terms was discussed in Ref. \cite{WF99}. Here \( \phi 
_{1} \)
and \( \phi _{2} \) are evaluated at point \( x \), whereas \( \phi 
_{3} \)
and \( \phi _{4} \) are evaluated at point \( x' \). In the 
coincidence limit,
\( x'\rightarrow x \), the fully normal-ordered term is finite, but 
the cross
term and vacuum terms diverge. The singularity of the cross term is 
of particular
significance because, unlike the vacuum term, it is state-dependent. 
The fully
normal-ordered term will not contribute to the fluctuations so long 
as the quantum
state is a coherent state. The pure vacuum term will also not 
contribute so
long as we restrict our attention to the differences between a given 
quantum
state and the vacuum state.

If the quantum state is other than a coherent state, there are also 
state-dependent
stress tensor fluctuations in the fully normal ordered term. These 
fluctuations
were discussed in Refs.~\cite{F82,DF88,Kuo,PH97}, especially in a 
context where
the stress tensor is the source of gravity. The normal ordered term is 
always
finite and does not present a divergence problem, in contrast to the 
cross term.
The latter term can only be made meaningful if one examines space or 
time integrated
quantities and has a prescription for defining the resulting 
integrals. We may
schematically express the expectation value of a product of stress 
tensor operators
as 
\begin{equation}
\label{eq:Tdecomp}
\langle :T_{\mu \nu }::T_{\rho \sigma }:\rangle =\langle :T_{\mu \nu 
}T_{\rho \sigma }:\rangle +\langle T_{\mu \nu }T_{\rho \sigma 
}\rangle _{cross}+\langle T_{\mu \nu }T_{\rho \sigma }\rangle _{0}\, ,
\end{equation}
where the three terms of the right hand side are, respectively, the
fully normal ordered term, the cross term, and the vacuum term.
For a single mode coherent state \( |z\rangle  \), 
\begin{equation}
\label{eq:zz}
\langle z|:T_{\mu \nu }T_{\rho \sigma }:|z\rangle =\langle z|:T_{\mu 
\nu }:|z\rangle \langle z|:T_{\rho \sigma }:|z\rangle \, .
\end{equation}
In such a state, the fluctuations of the stress tensor are described 
by quantities
of the form 
\begin{equation}
\langle \triangle T^{2}\rangle =\langle :T_{\mu \nu }:^{2}\rangle 
-\langle :T_{\mu \nu }:\rangle ^{2}=\langle :T_{\mu \nu }:^{2}\rangle 
_{cross}+\langle :T_{\mu \nu }:^{2}\rangle _{0}\, .
\end{equation}
If we are interested only in the changes in \( \langle \triangle 
T^{2}\rangle  \)
when the quantum state is varied, then the pure vacuum term can be 
ignored,
and only the cross term is important: 
\begin{equation}
\langle \triangle T^{2}\rangle \rightarrow \langle :T_{\mu \nu 
}:^{2}\rangle _{cross}\, .
\end{equation}

Note that we do not mean to suggest that there is no physical meaning 
to the
pure vacuum term. It presumably describes fluctuations of the stress 
tensor
components in the Minkowski vacuum state. More precisely, if one 
measures a
spacetime averaged component, the result of the measurement should 
undergo fluctuations
which vary as an inverse power of the size of the averaging region. 
However,
in a non-vacuum state, the magnitude of the cross term will grow as 
the mean
energy density in the state. Thus there will be a regime in which the 
effects
of the cross term dominate those of the vacuum term. 

We must resolve the issue of the state-dependent divergences in the cross
term if it is to have any physical content. This issue was discussed
by us in Ref.~\cite{WF99}, where it was shown that although the stress tensor
correlation function is singular in the coincidence limit, integrals of
this function over space and time  can still be well-defined. The cross
term in the stress tensor correlation function has the form
\begin{equation}
\langle T_{\mu \nu }(x) T_{\rho \sigma}(x')\rangle _{cross}
= \frac{F(x,x')}{(x-x')^4} \, ,
\end{equation}
where $F(x,x')$ is a regular function of the spacetime points $x$ and $x'$,
and $(x-x')^2$ is the squared geodesic distance between them. Integrals of 
the correlation function appear to be formally divergent, but nonetheless
may be defined by an integration by parts procedure. Suppose for the sake
of illustration that the integrations are over time only, and note that
\begin{equation}
\frac{1}{(t-t')^4} = -\frac{1}{12}\, 
\frac{\partial^2}{\partial t^2} \frac{\partial^2}{\partial t'^2} 
\ln (t-t')^2 \,.
\end{equation}
If the function $F$ vanishes sufficiently rapidly at the endpoints of the
integrations, we can write
\begin{equation}
\int dt \, dt' \frac{F(t,t')}{(t-t')^4} = -\frac{1}{12}\,
\int dt \, dt' \ln (t-t')^2 \, 
\frac{\partial^2}{\partial t^2} \frac{\partial^2}{\partial t'^2} \,F(t,t') \,.
\end{equation}
This procedure provides a way to define integrals with singular integrands,
and has been discussed by various authors \cite{Davies,FJL}.
 As we will see below in Sect.~\ref{sec:STF} and in the Appendix, the integrals
of the cross term which describe radiation pressure fluctuations can be made 
finite by a similar procedure.

\section{Induced Momentum Fluctuations of a Single Mirror }
\label{sec: QFRP}

It is well known from classical physics that a beam of light falling 
on a reflecting
or absorbing surface exerts a pressure. This pressure may be computed 
by integration
of the appropriate component of the Maxwell stress tensor over the 
surface.
It may also be computed by counting photon momenta. Let us illustrate 
the latter
method, which we will call the ``photon number'' approach. If an 
incident
monochromatic beam of angular frequency \( \omega  \) and energy 
density \( \rho  \)
strikes a surface, the mean number of photons striking per unit time 
per unit
area is\footnote{
Units in which \( \hbar =c=1 \) will be used throughout this paper. 
Electromagnetic quantities are in Lorentz-Heaviside units.}
 \( \rho /\omega  \). If the light is perfectly reflected, each 
photon imparts
a momentum \( 2\omega  \) to the surface, resulting in a radiation 
pressure
of \( 2\rho  \). As expected, both the stress tensor and the photon 
number
approaches yield the same answer.

However, these calculation give only a mean value. The radiation 
pressure should
undergo fluctuations about this mean. In the photon number viewpoint, 
these
fluctuations arise from fluctuations in the rate of photons striking 
the surface.
In the stress tensor viewpoint, the fluctuations arise because the 
quantum state
of the radiation field is not an eigenstate of pressure. The main 
purpose of
this section is to examine radiation pressure fluctuations in a single
mode coherent state from both viewpoints, and to compare the results.
In one subsection, Sect.~\ref{sec: NES}, we will also examine the case
of a single mode photon number eigenstate.

\subsection{The Photon Number Approach \label{sec: PCV}}

In this approach, the radiation pressure fluctuates because of 
statistical fluctuations
in the numbers of photons striking the surface. Suppose that a beam 
of light
with angular frequency \( \omega  \) is described by a single mode coherent 
state, \( |z\rangle  \),
an eigenstate of the annihilation operator, \( a|z\rangle 
=z|z\rangle  \).
The mean number of photons which strike a mirror in time \( \tau  \) 
is 
\begin{equation}
\langle n\rangle =\langle a^{\dagger }a\rangle =|z|^{2}\, .
\end{equation}
If the mirror is perfectly reflecting, then the mean momentum 
transferred is
the expectation value of the operator 
\begin{equation}
\label{eq:momop}
p=2\omega n\, .
\end{equation}
The dispersion of this momentum is given by 
\begin{equation}
\label{eq:momdi}
\langle \Delta p^{2}\rangle =\langle p^{2}\rangle -\langle p\rangle 
^{2}=4\omega ^{2}(\langle n^{2}\rangle -\langle n\rangle ^{2})\, .
\end{equation}
In a coherent state, 
\begin{equation}
\langle n^{2}\rangle -\langle n\rangle ^{2}=\langle a^{\dagger 
}aa^{\dagger }a\rangle -\langle a^{\dagger }a\rangle ^{2}=\langle 
(a^{\dagger })^{2}a^{2}\rangle +\langle a^{\dagger }a\rangle -\langle 
a^{\dagger }a\rangle ^{2}=\langle a^{\dagger }a\rangle =\langle 
n\rangle \, .
\end{equation}
Thus 
\begin{equation}
\langle \Delta p^{2}\rangle =4\omega ^{2}\langle n\rangle =4\omega \, 
A\rho \, \tau \, ,   \label{eq:delP_pn}
\end{equation}
where \( \rho  \) is the mean energy density of the incident beam, 
and \( A \)
is its cross sectional area. If the mirror is a free body with mass 
\( m \),
the mean squared velocity fluctuation is 
\begin{equation}
    \label{eq:dv2_{pn}}
\langle \Delta v^{2}\rangle =4\frac{\omega A\rho }{m^{2}}\tau \, .
\end{equation}

\subsection{The Stress Tensor Approach }
\label{sec:STF}

An alternative approach to the problem of radiation pressure 
fluctuations is
the method of stress tensor fluctuations. It is well known that one 
can calculate
the force on a surface by integration of the relevant component of 
the stress
tensor over that surface. It thus seems reasonable to expect that the 
fluctuations
in this force can also be computed from the quantum stress tensor. 
There is,
however, a problem which needs to be resolved in this approach. This 
is that
products of stress tensor operators are not well defined at 
coincident points. Even the integrals of these products are formally 
divergent and need a regularization. A regularization scheme was used in 
our previous work \cite{WF99} and will be adopted again in this 
paper. The main idea is to treat the measuring process as a 
switch-on-switch-off process and do an integrations by parts. 

Consider a mirror
of mass \( m \) which is oriented perpendicularly to the \( x 
\)-direction.
If the mirror is at rest at time \( t=0 \), then at time \( t=\tau  
\) its
velocity in the \( x \)-direction is given classically by 
\begin{equation}
\label{eq:v}
v=\frac{1}{m}\int _{0}^{\tau }dt\int _{A}da\, \, T_{xx}\, ,
\end{equation}
where \( T_{ij} \) is the Maxwell stress tensor, and \( \int _{A}da 
\) denotes
an integration over the surface of the mirror. Here we assume that 
there is
radiation present on one side of the mirror only. Otherwise, 
Eq.~(\ref{eq:v})
would involve a difference in \( T_{xx} \) across the mirror. When 
the radiation
field is quantized, \( T_{ij} \) is replaced by the normal ordered 
operator
\( :T_{ij}: \), and Eq.~(\ref{eq:v}) becomes a Langevin equation. The 
dispersion
in the mirror's velocity becomes
\begin{equation}
\label{eq:v2_{0}}
\langle \triangle v^{2}\rangle =\frac{1}{m^{2}}\int _{0}^{\tau }dt\, 
\int _{0}^{\tau }dt'\, \int _{A}da\, \int _{A}da'\, [\langle 
:T_{xx}(x)::T_{xx}(x'):\rangle -\langle 
:T_{xx}(x):\rangle \langle :T_{xx}(x'):\rangle ]
\end{equation}
 As discussed above, when the quantum state of the radiation field is 
a coherent
state and we ignore the pure vacuum term, then the dispersion in \( 
T_{xx} \)
is given by the cross term alone, and 
\begin{equation}
\label{eq:v2}
\langle \triangle v^{2}\rangle =\frac{1}{m^{2}}\int _{0}^{\tau 
}dt\int _{0}^{\tau }dt'\int _{A}da\int _{A}da'\langle 
T_{xx}(x)T_{xx}(x')\rangle _{cross}\, .
\end{equation}
The components of the energy-momentum tensor for the electromagnetic 
field are
(Lorentz-Heaviside units are used here.) 
\begin{eqnarray}
T^{ij} & = & \delta ^{ij}T^{00}-(E^{i}E^{j}+B^{i}B^{j})\, 
,\label{eq:Tij} \\
T^{0i} & = & \epsilon ^{ijk}E^{j}B^{k}\, ,\label{eq:T0i} 
\end{eqnarray}
and 
\begin{equation}
\label{eq:T00}
T^{00}=\frac{1}{2}(E^{2}+B^{2})\, .
\end{equation}
Here \( E^{i} \) and \( B^{i} \) are Cartesian components of the 
electric
and magnetic fields, respectively. In particular, 
\begin{equation}
\label{eq:Txx}
T_{xx}=\frac{1}{2}(E_{y}^{2}+E_{z}^{2}+B_{y}^{2}+B_{z}^{2})\, .
\end{equation}

We now assume that a linearly polarized plane wave is normally 
incident and
is perfectly reflected by the mirror. Take the polarization vector to 
be in
the \( y \)-direction, so that \( E_{z}=B_{y}=0 \). At the location 
of the
mirror, \( E_{y}=0 \), and only \( B_{z} \) contributes to the stress 
tensor.
Thus, when we apply Eq. (\ref{eq:phiprod}) to find \( \langle 
T_{xx}(x)T_{xx}(x')\rangle _{cross} \),
the only nonzero quadratic normal-ordered product will be \( \langle 
:B_{z}(x)B_{z}(x'):\rangle  \).
The result is 
\begin{equation}
\label{eq:Tcross2}
\langle T_{xx}(x)T_{xx}(x')\rangle _{cross}=\langle 
:B_{z}(x)B_{z}(x'):\rangle \langle B_{z}(x)B_{z}(x')\rangle _{0}\, .
\end{equation}
The \( z \)-component of the magnetic field operator may be expressed 
in terms
of mode functions as 
\begin{equation}
\label{eq:Bfield}
B_{z}(x)=\sum _{\omega }(a_{\omega }\, B_{\omega }+a_{\omega 
}^{\dagger }\, B_{\omega }^{*})\, ,
\end{equation}
where the mode function is 
\begin{equation}
\label{eq:Bmode}
B_{\omega }=-2\, C\cos (\omega x) \,e^{-i\omega t}\, . \label{eq:mode}
\end{equation}
Here C is the coefficient for box normalization in a volume \( V \)
\begin{equation}
\label{eq:C}
C=\sqrt{\frac{\omega }{2V}}\, .
\end{equation}
The coherent state is an eigenstate of the annihilation operator 
\begin{equation}
a_{\omega' }|z\rangle =\delta _{\omega' \omega }z|z\rangle \, ,
\end{equation}
where \( z \) is a complex number 
\begin{equation}
z=|z|e^{-i\phi }\, .
\end{equation}
The expectation value of the normal ordered product of field 
operators is now

\begin{equation}
\label{eq:normprod_{B}}
\langle z|:B_{z}(t_{1})B_{z}(t_{2}):|z\rangle =16C^{2}|z|^{2}\cos 
(\omega t_{1}+\phi )\cos (\omega t_{2}+\phi )\cos (\omega x_{1})\cos 
(\omega x_{2})
\end{equation}
The vacuum magnetic field two-point function in the presence of a 
perfectly
reflecting plane at \( z=0 \) is given by 
\begin{equation}
\label{eq:2pt}
\langle B_{z}(t_{1},{\textbf {x}}_{1})B_{z}(t_{2},{\textbf 
{x}}_{2})\rangle _{0}=\langle B_{z}(t_{1},{\textbf 
{x}}_{1})B_{z}(t_{2},{\textbf {x}}_{2})\rangle _{E0}+\langle 
B_{z}(t_{1},{\textbf {x}}_{1})B_{z}(t_{2},{\textbf {x}}_{2})\rangle 
_{I0}\, .
\end{equation}
The first term is the two-point function for empty space, 
\begin{equation}
\label{eq:2pt_{e}mpty}
\langle B_{z}(t_{1},{\textbf {x}}_{1})B_{z}(t_{2},{\textbf 
{x}}_{2})\rangle _{E0}=\frac{(t_{1}-t_{2})^{2}+|{\textbf 
{x}}_{1}-{\textbf {x}}_{2}|^{2}-2(z_{1}-z_{2})^{2}}{\pi 
^{2}[(t_{1}-t_{2})^{2}-|{\textbf {x}}_{1}-{\textbf 
{x}}_{2}|^{2}]^{3}}\, .
\end{equation}
The second term is an image term 
\begin{equation}
\langle B_{z}(t_{1},{\textbf {x}}_{1})B_{z}(t_{2},{\textbf 
{x}}_{2})\rangle _{I0}=\langle B_{z}(t_{1},{\textbf 
{x}}_{1})B_{z}(t_{2},{\textbf {x}}_{2})\rangle _{E0}\biggl 
|_{z_{2}\rightarrow -z_{2}}\, .
\end{equation}
Both terms give equal contributions to the radiation pressure 
fluctuations on
a mirror located at \( z=0 \).

We can now combine these results to write Eq.~(\ref{eq:v2}) as
\begin{equation}
\langle \triangle v^{2}\rangle =\frac{32C^{2}|z|^{2}}{\pi 
^{2}m^{2}}\int _{A}da_{1}\, \int da_{2}\, J
\end{equation}
where 
\begin{equation}
\label{eq:J}
J=\int _{0}^{\tau }dt_{1}\int _{0}^{\tau 
}dt_{2}\frac{(t_{1}-t_{2})^{2}-a}{[(t_{1}-t_{2})^{2}-b^{2}]^3}\, \cos 
(\omega t_{1})\cos (\omega t_{2})\, ,
\end{equation}
with 
\begin{equation}
a=(z_{1}-z_{2})^{2}-(y_{1}-y_{2})^{2}
\end{equation}
and 
\begin{equation}
b^{2}=(y_{1}-y_{2})^{2}+(z_{1}-z_{2})^{2}\, .
\end{equation}
Here we have set \( \phi = \arg(z) =0 \), as it just shifts the origin of time, 
and taken
the location of the mirror to be at \( x=0 \). The integral \( J \) 
is evaluated
in the Appendix in the limit of large \( \tau  \), with the result 
\begin{equation}
\label{eq:J2}
J=\frac{2\pi \tau }{32b^{5}}\, \left\{ \left[ b^{2}(b^{2}+a)\omega 
^{2}+b^{2}-3a\right] \sin b\omega +\omega b(3a-b^{2})\cos b\omega 
\right\} \, .
\end{equation}
The singularities in the integrand of  Eq.~(\ref{eq:J}) are third order poles,
which are evaluated using an integration by parts.

We next need to perform the spatial integration over the area of the 
mirror
which is illuminated by the laser beam. Assume that the illuminated 
region is
a disk of radius \( R \) and hence area \( A=\pi R^{2} \), and that 
the incident flux is uniform over this disk. (This assumption is not essential,
but simplifies the calculations.) If we take the origin for the \( 
da_{1} \)
integration to be at \( y_{2}=z_{2}=0 \), then 
\begin{eqnarray}
I & \equiv  & \int _{A}da_{2}\int _{A}da_{1}\, [b^{2}(b^{2}+a)\omega 
^{2}+b^{2}-3a]\sin b\omega +\omega b(3a-b^{2})\cos b\omega ]\, 
b^{-5}\nonumber \\
 & = & \int da_{2}\int _{0}^{R}r\, dr\int _{0}^{2\pi }d\theta \, 
\frac{1}{r^{5}}\nonumber \\
 &  & \times \left[ \omega r(r^{2}-3r^{2}\sin ^{2}\theta )\cos 
(\omega r)+[3r^{2}\sin ^{2}\theta +r^{4}\omega ^{2}-r^{2}(1+r^{2}\sin 
^{2}\theta \omega ^{2})]\sin (\omega r)\right] \nonumber \\
 & = & \pi \int da_{2}\int _{0}^{R}\frac{(1+\omega ^{2}r^{2})\sin 
(\omega r)-\omega r\cos (\omega r)}{r^{2}}\, dr\, .
\end{eqnarray}
Further assume that \( R\gg \omega ^{-1} \). Then we can let \( 
R\rightarrow \infty  \)
in the upper limit of the \( r \) integration. The $d a_2$ integration
now simply contributes a faxtor of $A$, and we have  
\begin{eqnarray}
I & \approx  & \pi A\int _{0}^{\infty }\frac{(1+\omega ^{2}r^{2})\sin 
(\omega r)-\omega r\cos (\omega r)}{r^{2}}\, dr\nonumber \\
 & = & \pi \omega A\int _{0}^{\infty }\frac{(1+u^{2})\sin (u)-u\cos 
(u)}{u^{2}}\, du\nonumber \\
 & = & \pi \omega A\int _{0}^{\infty }\, du\, \left[ \sin 
u-\frac{d}{du}\left( \frac{\sin u}{u}\right) \right] \nonumber \\
 & = & 2\pi \omega A\, .
\end{eqnarray}
In the last step, we used \( \int _{0}^{\infty }\, du\, \sin u=\lim 
_{\alpha \rightarrow 0}\int _{0}^{\infty }\, du\, \sin u\, 
{\textrm{e}}^{-\alpha u}=1\, . \)
Thus we have 
\begin{equation}
\langle \triangle v^{2}\rangle =\frac{16\, C^{2}|z|^{2}\omega A\tau 
}{\pi m^{2}}\, .
\end{equation}
The energy density in the incident wave can be written as 
\begin{equation}
\rho =\frac{\omega |z|^{2}}{V}=2C^{2}|z|^{2}\, ,
\end{equation}
so we can express the velocity fluctuations as 
\begin{equation}
\label{eq:dv2_{s}t}
\langle \triangle v^{2}\rangle =4\, \frac{A\omega \rho }{m^{2}}\, 
\tau \, .
\end{equation}
Note that this result agrees with that from the photon counting 
approach, Eq.~(\ref{eq:dv2_{pn}}).

\subsection{The Wavepacket Approach \label{sec: WA}}

Here we wish to provide an alternative derivation of the  momentum 
fluctuations of a single mirror
using the stress tensor approach. Rather than performing all
of the calculations in coordinate space, as was done in  Sect.~\ref{sec:STF},
we will use an approach based upon wavepacket modes. This approach will
prove useful in discussing interferometer noise.
Assume that the occupied mode is a 
wavepacket which is sharply peaked at frequency \( \omega \). 
Using  Eq.~(\ref{eq:v2}) and Eq.~(\ref{eq:Tcross2})
for a coherent state, the momentum fluctuation becomes
\begin{eqnarray}
\langle \triangle p^{2}\rangle  & = & m^{2}\langle \triangle 
v^{2}\rangle \nonumber \\
 & = & \int dt\, da\, \int dt'\, da'\, \langle 
T_{xx}(x)T_{xx}(x')\rangle _{cross}\nonumber \\
 & = & \int dt\, da\, \int dt'\, da'\langle 
:B_{z}(x)B_{z}(x'):\rangle \langle B_{z}(x)B_{z}(x')\rangle _{0}\, 
.\label{eq:dp^2} 
\end{eqnarray}
Let the magnetic field operator \( B_{z}(x) \) be expanded in terms 
of a complete set of positive
frequency wavepacket modes \( \{u_{j}(x)\} \) :
\begin{equation}
\label{eq:b-mode}
B_{z}(x)=\sum _{j}[a_{j}\, u_{j}(x)+a_{j}^{\dagger }u_{j}^{*}(x)]\, .
\end{equation}
 For our purpose, we take these modes to be fairly sharply peaked in 
frequency
and use the normalization condition
\begin{equation}
\int u_{j}^{*}u_{j'}\, d^{3}x=\frac{1}{2}\omega _{j}\, \delta_{jj'} \, ,
\end{equation}
where \( \omega _{j} \) is the mean frequency of packet \( j \). More 
generally,
we should expand the vector potential \( {\bf A} \) in terms of wavepackets \( 
f_{j} \):
\[
{\bf A}=\sum _{j}(a_{j}\widehat{e_{j}}f_{j}+H.c.)\, ,
\]
where 
\( \langle f_{j},f_{j'}\rangle =\delta _{jj'} \) 
and 
\( \langle f_{j},f_{j'}\rangle  \)
is the Klein-Gordon inner product. Then the modes \( u_{j} \) are 
expressed
as derivatives of the modes \( f_{j} \). Consider a single mode 
coherent state
\( |z\rangle  \) as the quantum state, and let the mode be a 
wavepacket \( u_{0} \).
Then with a suitable choice of the phase of this mode function, we 
can write
\begin{equation}
\label{eq:normal-b}
\langle :B_{z}(x)B_{z}(x'):\rangle 
=|z|^{2}[u_{0}(x)+u_{0}^{*}(x)][u_{0}(x')+u_{0}^{*}(x')]\, .
\end{equation}
Note that the integrals in \( \langle \triangle p^{2}\rangle  \) are 
of the
form \( \int dt\, da \) rather than \( \int d^{3}x \) . If the 
integrand is
a function of \( t-x \) alone, or \( t+x \) alone, these are 
equivalent:
\begin{equation}
\int _{-\infty }^{\infty }dt\, f(t-x)=\int _{-\infty }^{\infty }du\, 
f(u)=\int _{-\infty }^{\infty }du\, f(-u)=\int _{-\infty }^{\infty 
}dx\, f(t-x)
\end{equation}
and
\begin{equation}
\int _{-\infty }^{\infty }dt\, f(t+x)=\int _{-\infty }^{\infty }dx\, 
f(t+x)\, ,
\end{equation}
where \( u=t-x \). However, when the mirror is present, \( u_{0} \) 
contains pieces moving in both directions: 
\begin{equation}
\label{eq:ui+ur}
u_{0}=u_{0I}+u_{0R\, ,}
\end{equation}
where \( u_{0I} \) is the incident wavepacket and \( u_{0R} \) is the 
reflected
wavepacket. The key feature that we will use is that \( u_{0} \) is 
orthogonal
to \( u_{j} \) (\( j\neq 0 \)), but \( u_{0I} \) and \( u_{0R} \) are 
not
orthogonal to each other. Inserting Eq. (\ref{eq:b-mode}) and Eq. 
(\ref{eq:normal-b})
into Eq. (\ref{eq:dp^2}) yields
\begin{eqnarray}
    \label{eq:dp2_{u}}
\langle \triangle p^{2}\rangle  & = & |z|^{2}\int dt\, da\, \int dt'\, 
da'[u_{0}(x)+u_{0}^{*}(x)][u_{0}(x')+u_{0}^{*}(x')]\nonumber \\
 &  & \times \frac{1}{2}\sum 
_{j}[u_{j}(x)u_{j}^{*}(x')+u_{j}^{*}(x)u_{j}(x')]\nonumber \\
 & = & |z|^{2}\sum _{j}Re\left[\int dt\, da\, \int dt'\, 
da'u_{0}(x)u_{0}^{*}(x')u_{j}^{*}(x)u_{j}(x')\right]\nonumber \\
 & = & |z|^{2}\sum _{j}Re\left[\int u_{0}(x)u_{j}^{*}(x)dt\, da'\, \int 
u_{0}^{*}(x')u_{j}(x')dt'\, da\right]\nonumber \\
 & = & |z|^{2}\left[\int u_{0}(x)u_{0}^{*}(x)dt\, da \right]^{2}\, . 
\end{eqnarray}
Using  Eq. (\ref{eq:ui+ur}), the integral in the last expression becomes
\begin{eqnarray}
I & = & \int u_{0}(x)u_{0}^{*}(x)dt\, da\nonumber \\
 & = & \int dt\, da\, (u_{0I}+u_{0R})(u_{0I}^{*}+u_{0R}^{*})\nonumber 
\\
 & = & \int dt\, da\, 
(|u_{0I}|^{2}+|u_{0R}|^{2}+u_{0I}u_{0R}^{*}+u_{0I}^{*}u_{0R})\, 
.\label{eq:u_0_int} 
\end{eqnarray}
Here the only difference between \( u_{0I} \) and \( u_{0R} \) is their 
direction of travel. If $u_{0I} = f(t-x)$, then $u_{0R} = f(t+x)$,
and $u_{0I} = u_{0R}$ at the mirror, and
\begin{equation}
\label{eq:uj=ur}
\int dt\, da\, u_{0I}u_{0R}^{*}=\int dt\, da\, u_{0I}^{*}u_{0R}=\int 
dt\, da\, |u_{0I}|^{2}=\int dt\, da\, |u_{0R}|^{2}\, .
\end{equation}
Thus 
\[
I = \int u_{0}(x)u_{0}^{*}(x)dt\, da=4\int dt\, da\, |u_{0I}|^{2}=4\int 
d^{3}x|u_{0I}|^{2}=2\omega \, ,\]
and the momentum fluctuation becomes
\begin{equation}
\langle \triangle p^{2}\rangle =4\omega^{2} |z|^{2}=4\omega^{2} \langle 
n\rangle \, . \label{eq:p2one}
\end{equation}
This is same as the result of the photon number counting approach,
Eqs.~(\ref{eq:dv2_{pn}}), and of the integration
by parts in Sect. \ref{sec:STF}, (\ref{eq:dv2_{s}t}).

\subsection{A Number Eigenstate \label{sec: NES}}

Most of this paper deals with single mode coherent states. However, in
this subsection, we wish to turn aside from the main line of development
and discuss the case of a single mode number eigenstate. It is apparent
in the photon number approach that there should not be any radiation
pressure fluctuations in such a state. In the stress tensor approach, 
the situation is less clear, as both the fully normal-ordered term
and the cross term are nonzero. 

We assume that the quantum state \( |n\rangle  \) is a number eigenstate
of a single mode.
Expand the magnetic field operator in terms of complete set of mode 
functions using
Eq.~(\ref{eq:Bfield}) to find the expectation value of
the fully normal ordered term
\begin{eqnarray}
 &  & \langle n|:T_{xx}(x)T_{xx}(x'):|n\rangle \nonumber \\
 & = & \frac{1}{4}\langle n|:B_{z}^{2}(x)B_{z}^{2}(x'):|n\rangle 
\nonumber \\
 & = & \frac{1}{4}n(n-1)[B_{\omega }^{*2}(x)B_{\omega 
}^{2}(x')+B_{\omega }^{2}(x)B_{\omega }^{*2}(x')+4|B_{\omega 
}(x)|^{2}|B_{\omega }(x')|^{2}]\, .
\end{eqnarray}
Here $B_{\omega }$ is the mode function for the occupied mode, and is assumed 
to be given by Eqs.~(\ref{eq:Bmode}) and (\ref{eq:C}).
A similar procedure leads to the result for the cross term
\begin{eqnarray}
 &  & \langle n|:T_{xx}(x)::T_{xx}(x'):|n\rangle _{cross}\nonumber \\
 & = & \langle n|:B_{z}(x)B_{z}(x'):|n\rangle \langle 
B_{z}(x)B_{z}(x')\rangle _{0}\nonumber \\
 & = & n \left[B_{\omega }(x)B_{\omega }^{*}(x')+B_{\omega 
}^{*}(x)B_{\omega }(x')\right] \langle B_{z}(x)B_{z}(x')\rangle _{0} \,.
\end{eqnarray}
The mean pressure is 
\begin{equation}
\langle :T_{xx}(x):\rangle_n  = \langle n|:T_{xx}(x):|n\rangle 
    =n|B_{\omega }(x)|^{2}\, .
\end{equation}

The momentum deviation due to the fully normal ordered term becomes
\begin{eqnarray}
\langle n|:\triangle p^{2}:|n\rangle _{n} & = & \langle :p^{2}:\rangle 
_{n}-\langle :p:\rangle _{n}^{2}\nonumber \\
 & = & \int dt\, da\, \int dt'\, da'\, (\langle 
:T_{xx}(x)T_{xx}(x'):\rangle _{n}-\langle :T_{xx}(x):\rangle 
_{n}\langle :T_{xx}(x'):\rangle _{n})\nonumber \\
 & = & \frac{n(n-1)}{4}\int dt\, da\, \int dt'\, da'\, (B_{\omega 
}^{*2}(x)B_{\omega }^{2}(x')+B_{\omega }^{2}(x)B_{\omega 
}^{*2}(x'))\nonumber \\
 &  & -n\, \int dt\, da\, \int dt'\, da'\, |B_{\omega 
}(x)|^{2}|B_{\omega }(x')|^{2}\, .\label{eq:-n} 
\end{eqnarray}
Integrals such as $\int dt\,B_{\omega }^{2}(x)$ contain rapidly oscillating 
integrands. We will assume that these integrals average to zero and can
be ignored. The remaining integrals, involving $|B_{\omega }|^{2}$ are
straight forward. The result for the contribution of the fully normal ordered 
term to the momentum deviation is
\[
\langle :\triangle p^{2}:\rangle _{n}=-4n\, \omega ^{2}\, .
\]
The cross term is
\begin{eqnarray}
\langle n|\triangle p^{2}|n\rangle _{cross} & = & \int dt\, da\, \int 
dt'\, da'\, \langle n|:B_{z}(x)B_{z}(x'):|n\rangle \langle 
B_{z}(x)B_{z}(x')\rangle _{0}\nonumber \\
 & = & n|C|^{2}\int dt\, da\, \int dt'\, da'\, \cos\, \omega x\, \cos\, 
\omega x'\langle B_{z}(x)B_{z}(x')\rangle _{0}\nonumber \\
 & = & 4n\omega ^{2}\, .\label{eq:n} 
\end{eqnarray}
Note that the cross term calculation is almost identical to the 
calculation
in the case of a coherent state. The momentum deviation due to the 
fully normal
ordered term and the cross term cancel out each other and yield zero 
momentum deviation
\begin{equation}
\langle n|\triangle p^{2}|n\rangle =\langle n|:\triangle 
p^{2}:|n\rangle +\langle n| \triangle p^{2}|n\rangle _{cross}=0\, ,
\end{equation}
which is expected in photon number approach. This calculation shows the
agreement between the photon number and the stress tensor approaches.
In the stress tensor approach, the fluctuations in the normal ordered
term are anticorrelated with those described by the cross term.

\section{Noise in an Interferometer }
\label{sec: Noise}

\begin{figure}
{\centering 
\leavevmode\resizebox*{!}{6.5cm}{\includegraphics{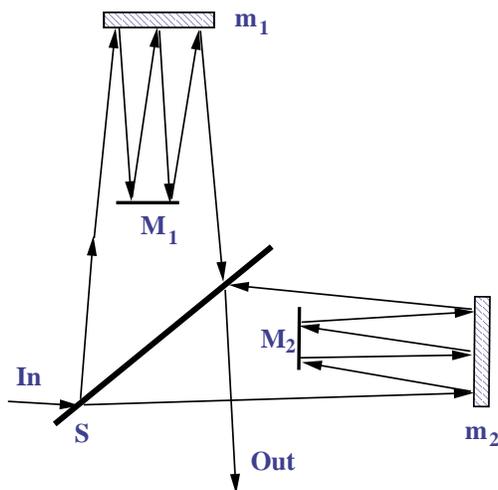}} \par}

\caption{The essentials of a Michelson interferometer. An input laser 
beam (``In'')
is split by the beam splitter S. The split beam bounces \protect\( 
b\protect \)
times (here \protect\( b=3\protect \)) in each arm between a free 
mass (\protect\( m_{1}\protect \)
or \protect\( m_{2}\protect \)) and a fixed mirror (\protect\( 
M_{1}\protect \)
or \protect\( M_{2}\protect \)) before being recombined to form the 
output
beam (``Out''). The mirrors of interest for the radiation pressure 
fluctuations
are located on the free masses. }

\label{fig:interferometer}
\end{figure}

A primary application of the result in Sect.~\ref{sec: PCV} is to 
estimate
the radiation pressure noise in an interferometer. 
(See Fig. \ref{fig:interferometer}) The laser beam
bounces \( b \) times in a arm of the interferometer before being 
recombined.
The masses at the end of each arm are subject to velocity and 
position uncertainty due to the radiation pressure fluctuations. \par

\subsection{Position Uncertainty in the Photon Number Approach}
\label{sec:pu_pn}

In this subsection, we will review the conventional, photon number
 approach to calculating interferometer noise \cite{Caves1,Caves2}.
The effect of multiple bounces is
accounted for by multiplying the momentum operator in Eq. (\ref{eq:momop})
by a factor 0f $b$
\begin{equation}
p=2b\omega n\, .
\end{equation}
This introduces a factor of \( b^{2} \) in the mean squared velocity 
fluctuation
\begin{equation}
\label{eq:v2d_{p}n}
\langle \Delta v^{2}\rangle =4b^{2}\frac{\omega A\rho }{m^{2}}\tau \, 
.
\end{equation}
The root mean squared position uncertainty of each mirror due to 
radiation pressure
fluctuations is then of order 
\begin{equation}
\label{eq:zrp}
{\Delta x}_{rp}=b\frac{\sqrt{\omega P}}{m}\, \tau ^{\frac{3}{2}}\, ,
\end{equation}
where \( P=A\rho  \) is the mean power in the laser beam. There is 
another
source of noise in the interferometer, the photon counting error, 
also known
as shot noise. This arises from the uncertainty in the location of an 
interference
fringe when a finite number of photons are counted, and is of order 
\begin{equation}
{\Delta x}_{pc}=\frac{1}{2b\sqrt{\omega P\tau }}\, .
\end{equation}
If we minimize the net squared position uncertainty 
\begin{equation}
({\Delta x}_{rp})^{2}+({\Delta x}_{pc})^{2}
\end{equation}
with respect to \( P \), the result is the optimum power 
\begin{equation}
\label{eq:Popt_{p}n}
P_{opt}=\frac{m}{2\omega \tau ^{2}b^{2}}\, .
\end{equation}
When \( P=P_{opt} \), the position uncertainty becomes 
\begin{equation}
\label{eq:sql}
\Delta x =\Delta x_{sql}=\sqrt{\frac{\tau }{m}}\, ,
\end{equation}
known as the \textit{standard quantum limit}.

The standard quantum limit is the position uncertainty which one 
obtains after
time \( \tau  \) by preparing a particle in a wavepacket state with 
initial
spatial width \( \Delta x_{0} \) and momentum spread \( \Delta 
p=m\Delta x_{0}/\tau  \).
After time \( \tau  \) has elapsed, the spread in the width of the 
wavepacket,
\( \Delta p\, \tau /m \), is of the order of \( \Delta x_{0} \). Thus 
the
standard quantum limit can be interpreted as being the minimum 
position uncertainty
which can be maintained for a time of the order of \( \tau  \).

\subsection{Position Uncertainty in the Stress Tensor Approach}

Here we wish to look in more detail at how the position uncertainty
arises in a description involving the quantum stress tensor.
 The position of each mass is disturbed by the pressure 
fluctuations, leading to an error in the measurement of the optical path 
length. The displacement of a given mass
is described by the time integral of Eq.~(\ref{eq:v})
\begin{equation}
\label{eq:x}
x-x(0)=\frac{1}{m}\int _{0}^{\tau }\int _{0}^{t}\int 
_{A}T_{xx}(x_{1})da_{1}dt_{1}dt
\end{equation}
The dispersion in position is then given by an expression analogous 
to Eq. (\ref{eq:v2_{0}})
\begin{eqnarray}
    \label{eq:x2}
\langle \triangle x^{2}\rangle  & \equiv  & \langle x^{2}\rangle 
-\langle x \rangle ^{2}\nonumber \\
 & = & \frac{1}{m^{2}}\int _{A}\int _{A}da_{1}da_{2}\int _{0}^{\tau 
}\int _{0}^{\tau }dt\, dt'\int _{0}^{t}\int _{0}^{t'}\left[\langle 
:T_{xx}(x_{1})::T_{xx}(x_{2}):\rangle \right.\nonumber \\
 & - & \left.
\langle :T_{xx}(x_{1}):\rangle \langle :T_{xx}(x_{2}):\rangle 
\right]dt_{1}dt_{2}\, . 
\end{eqnarray}

As before, only the cross term in the stress tensor fluctuations will 
contribute when the quantum state is a coherent state.
If we use this fact, and then take the second derivative with respect 
to \( \tau  \),
we can write 
\begin{eqnarray}
\frac{d^{2}}{d\tau ^{2}}\langle \triangle x^{2}\rangle  & = & 
\frac{2}{m^{2}}\left[ \int _{0}^{\tau }dt_{1}\int _{0}^{\tau 
}dt_{2}\int da_{1}\int da_{2}\langle 
T_{xx}(t_{1})T_{xx}(t_{2})\rangle _{cross}\right. \nonumber \\
 &  & \left. +\int _{0}^{\tau }dt'\int _{0}^{t'}dt_{2}\int da_{1}\int 
da_{2}\langle T_{xx}(\tau )T_{xx}(t_{2})\rangle _{cross}\right] \, 
.\label{eq:x2diff} 
\end{eqnarray}
Now we need to assume that the laser beam is switched on in the past 
and then
switched off in the future. This issue was discussed in 
Ref.~\cite{WF99}, where
integrations by parts were performed in order to deal with the 
singular behavior
of the cross term. The asymptotic condition insures that the surface 
terms arising
in the integrations all vanishes. In the present calculation, we 
require that
the normal ordered factors vanish at \( t=\tau  \), and hence the 
second term
on the right hand side of Eq. (\ref{eq:x2diff}) vanishes. The 
remaining term
is proportional to \( \langle \triangle v^{2}\rangle  \), so that
\begin{equation}
\frac{d^{2}}{d\tau ^{2}}\langle \triangle x^{2}\rangle =2\, \langle 
\triangle v^{2}\rangle \, .
\end{equation}
Thus if $\langle \triangle v^{2}\rangle \propto \tau$, then 
$\langle \triangle x^{2}\rangle \propto \tau^3$. This calculation is
the justiication for Eq.~(\ref{eq:zrp}).

There are two major issues to be studied in remainder of this section. 
One is to study the effects of multiple bounces within one arm of the 
interferometer, which will be done in the following subsection.
 The other is to study whether there are any correlations 
between the two arms. The discussion in Sect.~\ref{sec:pu_pn} implicitly
assumed the absence of correlations, and a justication for this assumption
was given by Caves \cite{Caves1}. In the present context, there is a
correlation term \( \langle 
\triangle x_{12}^{2}\rangle  \)
which takes the form of Eq.~(\ref{eq:x2}) with \( x_{1} \) on one 
mirror and \( x_{2} \) on the other mirror. We will argue below 
in Sect.~\ref{sec:corr} that this term
is negligible compared to $\langle \triangle x^{2}\rangle$ for each arm
separately.

\subsection{Multiple Bounces in One Arm}
\subsubsection{A Delay Line}

Now we wish to consider the situation where a laser beam bounces several
times between a pair of mirrors, as illustrated in Fig.~1. This arrangement
is sometimes called a ``delay line''.  Suppose that the 
beam is recycled $b$ times within a single interferometer arm. We have already
seen that in the photon number approach, the momentum fluctuation of
the end mirror, $\langle \triangle p^{2}\rangle$ is now proportional to $b^2$.
Specifically,
\begin{equation}
\langle \triangle p^{2}\rangle = b^2\, \langle \triangle p^{2}\rangle_1 \, ,
                                              \label{eq:b_bounces}
\end{equation}
where $\langle \triangle p^{2}\rangle_1$ is the single bounce result given
in Eq.~(\ref{eq:p2one}). One can understand this result in the following way:
if there is a fluctuation in the number of photon entering the interferometer 
arm, that fluctuation is maintained on each of the successive bounces. If
slightly more than the expected number of photons hit the mirror on the first 
bounce, the same excess will reappear on later bounces. One can picture the
same photons as simply recycling $b$ times. However, in the stress tensor
approach, it is less obvious how the $b^2$ factor will arise. This factor
requires that the stress tensor fluctuations at the different spots on the
mirror be exactly correlated with one another. 

We can understand this by returning to Eq.~(\ref{eq:dp2_{u}}). If the 
integrations run over all time and over the entire area of the mirror,
then they pick up contributions from all of the bounces. Let us first suppose
that the spots formed on succesive bounces overlap on the same region
of the mirror. In this case, the mode function $u_0$ is approximately periodic
for $b$ periods:
\begin{equation}
u_0(t,{\bf x}) \approx u_0(t+2 L n,{\bf x})\, , \quad n = 1, 2, \cdots, b-1 \,.
                                                   \label{eq:period}
\end{equation}
Let the first bounce occur at time $t=0$, and subsequent bounces at
$t= 2 L n$. More precisely, these are the mean times at which the wavepacket 
hits the mirror. Let $T$ be some time interval which is long compared to
the length of the wavepacket, but short compared to $2L$.
Equation~(\ref{eq:u_0_int}) becomes
\begin{equation}
\langle \triangle p^{2}\rangle = |z|^2 \, I^2 \, ,
\end{equation}
where
\begin{equation}
I = \int_{-\infty}^{\infty} dt \int da \,u_{0}(x)u_{0}^{*}(x) 
 = \sum_{n=0}^{b-1} \int_{2 L n -T}^{2 L n +T} dt \int da \,
     u_{0}(x)u_{0}^{*}(x)\, .                 \label{eq:Isum}
\end{equation}
Note that the time intervals which are ignored in going from the first form
to the second are ones in which $u_{0}=0$ on the mirror, that is, in between
bounces. However, the periodicity property, Eq.~(\ref{eq:period}), implies
that each term in the sum is equal. Furthermore, each term gives the same 
contribution to $I$ as was found in the single bounce case:
\begin{equation}
I = b \int_{ -T}^{T} dt \int da\,  u_{0}(x)u_{0}^{*}(x)\,
  = 2 b \omega \,.
\end{equation}
Thus we obtain Eq.~(\ref{eq:b_bounces}).

Note that it does not matter whether the spots formed on the various bounces
actually overlap on the mirror or not. If they do not, then 
Eq.~(\ref{eq:period}) is replaced by a more complicated relation involving
an offset in position for the different bounces. However, once the area
integration is performed, this is irrelevant, and we still obtain 
Eq.~(\ref{eq:Isum}). Note that we are assuming that on all bounces, the beam
is nearly perpendiular to the mirror.

\subsubsection{A Fabry-Perot Cavity}
\label{sec:Fabry}

\begin{figure}
{\centering 
\leavevmode\resizebox*{!}{6.5cm}{\includegraphics{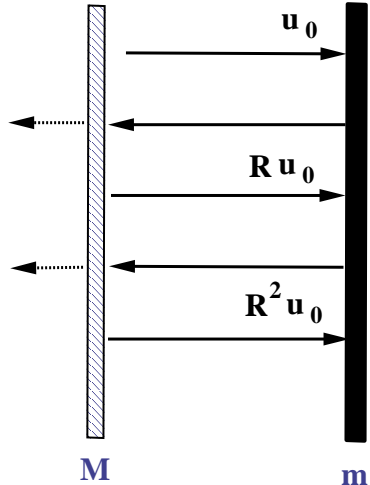}} \par}

\caption{A Fabry-Perot cavity. Here the right mirror $m$ is perfectly
reflecting, and the left mirror $M$ is partially reflecting, with a reflection 
amplitude $R$. An initial wavepacket $u_0$ returns to $m$ as $R u_0$ on the
first bounce, as $R^2 u_0$ on the second bounce, ect. }

\label{fig:Fabry}
\end{figure}

The delay line arrangement sketched in Fig.~\ref{fig:interferometer} 
and discussed above implies
a precise number of bounces. Another possibility, which is more likely to
be used in actual interferometers, is the Fabry-Perot cavity, illustrated
in Fig.~\ref{fig:Fabry}. Here at least one of the mirrors is 
partially reflecting, leading
to a finite storage time for a wavepacket in the cavity. We will discuss 
the case where the mirror on the free mass is assumed to be perfect, but
the opposite mirror in the cavity is not. This assumption allows us to 
continue to use our previous expressions, especially Eq.~(\ref{eq:dp^2}).
If the mirror on the free mass is not perfect, then it is necessary to
modify this expression and include electric field terms as well.

 Let $R$ be the complex reflection amplitude for the imperfect mirror,
so that $|R|^2$ is the fraction of the power reflected on each bounce. 
We assume that once inside the cavity, a given wavepacket mode bounces
an infinite number of times, but with diminishing amplitude. The 
effective number of bounces, $b'$, can be defined by
\begin{equation}
|R|^{2b'} = \frac{1}{2} \,.
\end{equation}
Thus the energy stored in an occupied wavepacket mode is reduced by a
factor of $\frac{1}{2}$ after $b'$ bounces.
We can then write
\begin{equation}
|R|^2 = e^{-\frac{\ln 2}{b'}} \approx 1 - \frac{\ln 2}{b'} \,, 
                                                  \quad b' \gg 1 \,.
\end{equation}

The effect of the finite reflectivity of the left mirror is to introduce a
factor of $r$ each time the wavepacket returns to the right mirror. Recall
that the magnetic field mode $u_0$ has no phase shift upon reflection from
the perfect mirror. We can express this as the following condition on
the mode function:
\begin{equation}
u_0(t+2 L n,{\bf x}) = R^n \,u_0(t,{\bf x}), \quad n= 0, 1, 2, \cdots,
\quad -T < t < T \,.
\end{equation}
That is, $u_0(t,{\bf x})$ for $-T < t < T$ is the initial form of the 
wavepacket when it hits the left mirror for the first time. The above relation
gives its form when it returns for the $n$-th time. 

We can now use this relation to write the analog of Eq.~(\ref{eq:Isum}):
\begin{equation}
I = \int_{-\infty}^{\infty} dt \int da\, u_{0}(x)u_{0}^{*}(x) 
 = \sum_{n=0}^{\infty} \int_{-T}^{T} dt \int da \,
    R^n u_{0}\, {R^*}^n(x)u_{0}^{*}(x)
 = \frac{1}{1-|R|^2}\, I_1 .                 \label{eq:Isum2}
\end{equation} 
We can now combine this with Eq.~(\ref{eq:dp2_{u}}) to write
\begin{equation}
\langle \triangle p^{2}\rangle = \left(\frac{b'}{\ln 2}\right)^2\, 
                        \langle \triangle p^{2}\rangle_1 \, ,
                                              \label{eq:Fabry}
\end{equation}
This result is the analog of  Eq.~(\ref{eq:b_bounces}) for the case of
a Fabry-Perot cavity. In both cases, the momentum fluctuations grow as 
the square of the effective number of bounces.

\subsection{The Equal Arm Interferometer}

In an equal arm interferometer with a perfect (loss-free) 50-50 beam splitter,
the input power is divided equally between the two arms. In the late
1970's, there was a controversy over whether radiation pressure fluctuations
will create noise in such an interferometer. The arguments reviewed at
the beginning of this section leading to the standard quantum limit,
Eq.~(\ref{eq:sql}), assume that the radiation pressure fluctuations in the
two arms are uncorrelated. However, one would expect that a fluctuation
which sends more power into one arm will cause a corresponding deficit 
in the other arm. This would lead to anticorrelated pressure fluctuations.
Caves \cite{Caves1,Caves2} resolved this controversy in the context of the
photon number approach. He showed that when vacuum modes which enter an
unused port of the interferometer are included, the fluctuations
are uncorrelated. In this subsection, we will rederive this result using
the stress tensor approach.

\subsubsection{Properties of a Beam Splitter}

\begin{figure}

{\centering 
\leavevmode\resizebox*{!}{11 cm}{\includegraphics{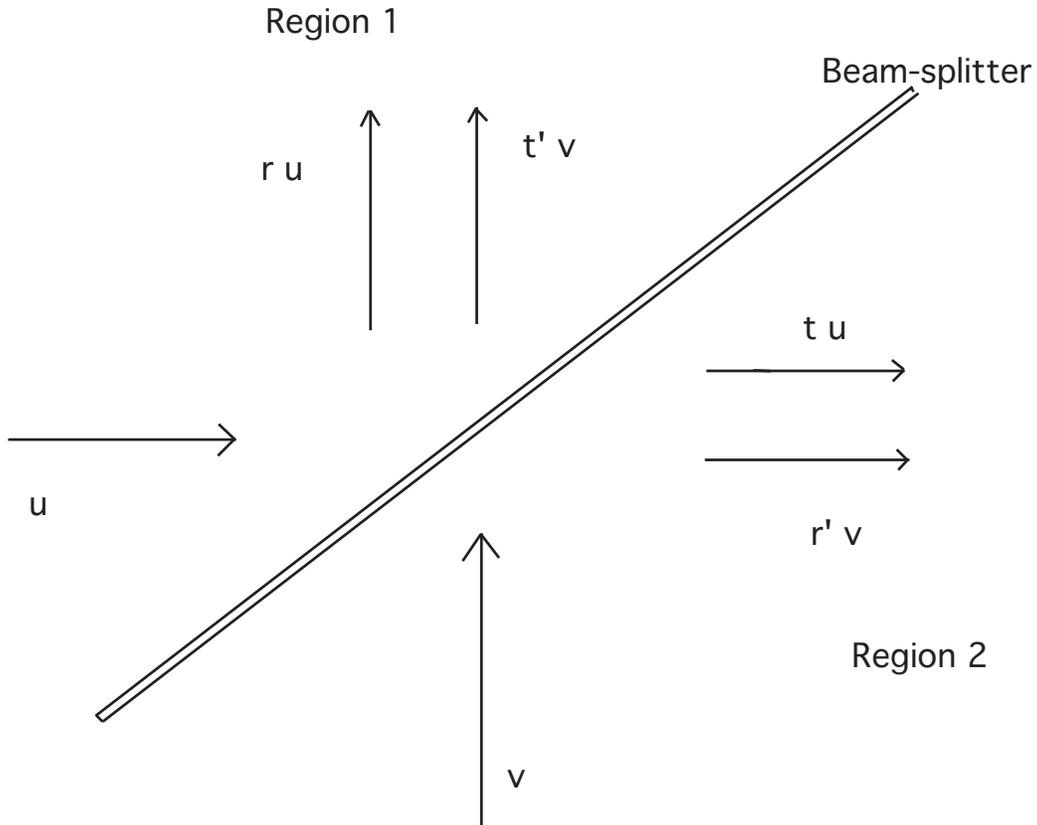}} \par}
\caption{ u and v represent the two incident wavepacket. \( r\, ,t \) are the 
complex amplitude reflectivity and transmissivity for incident light 
u, and  \( r'\, ,t' \) are for the incident light v} 
\label{fig:beam-splitter}
\end{figure}

Let us  consider a  perfect beam splitter. 
(See Fig. \ref{fig:beam-splitter}) It needs to 
satisfy the reciprocity relations, originally derived by Stokes in 1849,
\begin{equation}
    \label{eq:r=t}
|r|=|r'|\, \, ,\, \, |t|=|t'|\, ,
\end{equation}

\begin{equation}
\label{onewave}
|r|^{2}+|t|^{2}=1
\end{equation}
and
\begin{equation}
\label{twowave}
r'\, t^{*}+r^{*}t'=0\, ,
\end{equation}
where \( r\, ,t \) are the complex amplitude reflectivity and 
transmissivity
for light incident from one side and \( r'\, ,t' \) for light from 
the other
side. The first equation, Eq. (\ref{eq:r=t}), arises form the assumption 
that the reflectivity is the same from both sides.
The other two equations are the results due to the additional 
assumption of a
no-loss beam splitter. Assume that we send in a wavepacket \( u_{0} 
\) into
the beam splitter in the \( x \) direction. Then it will reflect the 
amount
\( r\, u_{0} \) to region \( 1 \) and transmit \( t\, u_{0} \) to 
region
\( 2 \). If this is a no-loss beam splitter, then the inverse 
operation will
bring the reflected and transmitted wavepackets back to the original 
incident
wavepacket, 
\begin{equation}
(r\, u_{0})r^{*}+(t\, u_{0})t^{*}=u_{0}\, ,
\end{equation}
which leads to Eq.~(\ref{onewave}). If there is another 
wavepacket \( u_{1} \) coming
into the beam splitter from the \( y \) direction as well as the one 
\( u_{0} \)
from \( x \), then in region \( 1 \) we have the transmitted 
wavepacket \( t'\, u_{1} \)
in additional to the reflected wavepacket \( r\, u_{0} \), Similarly 
we get
\( t\, u_{0}+r'\, u_{1} \) in the region \( 2 \). Again the inverse 
operation
 yields the relations
\begin{equation}
(r\, u_{0}+t'\, u_{1})r^{*}+(t\, u_{0}+r'\, u_{1})t^{*}=u_{0}
\end{equation}
and
\begin{equation}
(r\, u_{0}+t'\, u_{1})t'^{*}+(t\, u_{0}+r'\, u_{1})r'^{*}=u_{1}
\end{equation}
for a no-loss, perfect beam splitter. Both equations here will lead 
to the reciprocity
relation Eq.~(\ref{twowave}). For a \( 50-50 \) beam splitter, \( 
r'=r \),
\( t'=t \). Without losing generality, we may take the coefficients to 
be
\[
r=\frac{1}{\sqrt{2}}e^{i\, \phi _{r}\, }\]
and
\[
t=\frac{1}{\sqrt{2}}e^{i\, \phi _{t}\, ,}\]
where \( \phi _{r}\, ,\phi _{t} \) are the phase change due to 
reflection and transmission, respectively. 
 Plug them into the reciprocity relation 
Eq. (\ref{twowave}),
and the phase difference becomes
\begin{equation}
\label{phasediff}
\triangle =\phi _{r}-\phi _{t}=\frac{n\pi }{2}\, \, ,n=1,3,5...\, \, 
\, .
\end{equation}
This phase difference is crucial in the discussion of the momentum 
correlation of the two end mirrors in interferometer.

\subsubsection{Correlation between two arms}
\label{sec:corr}

In a quantum mechanical treatment of the radiation field in the presence of
a beam splitter, one often speaks of light entering one port of the
interferometer and of vacuum entering the other, unused port 
\cite{Caves1,Caves2,MF-optics}. From our point of view, this language is 
misleading. Vacuum modes are everywhere, and are entering both ports.
Furthermore, there are an infinite number of such vacuum modes, and one
would like to see more clearly which ones are actually relevant in a given
situation.

 Assume that wavepacket \( u_{0} \) (the occupied mode)
 with a particular frequency is incident on the beam splitter.
It will reflect \( r\, 
u_{0} \)
to mirror \( \#1 \) and transmit \( t\, u_{0} \) to mirror \( \#2 
\) (See Fig. \ref{fig:beam-splitter}). 
Similarly,
at mirror \( \#1 \) there are modes \( r\, u_{j} \) reflected from the 
vacuum fields
coming from the input port and \( t\, u_{k} \) transmitted from the 
vacuum
fields coming from the output port. At mirror \( \#2 \), we have 
vacuum modes \( t\, u_{j} \)
and \( r\, u_{k} \) in addition to the occupied mode, \( t\, u_{0} \). 
If we consider the momentum
difference transferred to the mirrors \( p=p_{1}-p_{2} \), then the 
deviation
becomes 
\begin{equation}
\langle \triangle p^{2}\rangle =\langle \triangle p_{1}^{2}\rangle 
+\langle \triangle p_{2}^{2}\rangle -2\langle \triangle 
p_{1}\triangle p_{2}\rangle \, .
\end{equation}
 Now in arm \( \#1 \) the incident wavepacket is \( r\, u_{0} \) and 
the complete
set of wavepackets from vacuum are \( (\sum _{j}r\, u_{j}+\sum 
_{k}t\, u_{k}) \).
Follow the reasoning leading to  Eq. (\ref{eq:dp2_{u}}) in the one 
arm case; the momentum dispersion of mirror \( \#1 \) becomes 
\begin{eqnarray}
 &  & \langle \triangle p_{1}^{2}\rangle \nonumber \\
 & = & |z|^{2}Re[\int dt\int da\, r\, u_{0}(\sum _{j}r\, u_{j}+\sum 
_{k}t\, u_{k})^{*}\int dt'\int da'\, r^{*}u^{*}_{0}(\sum _{j}r\, 
u_{j}+\sum _{k}t\, u_{k})]\nonumber \\
 & = & |z|^{2}\left|\int dt\int da\, 
u_{0}u_{0}^{*}\right|^{2}*\frac{1}{4}Re[(1+e^{i\triangle 
})(1+e^{-i\triangle })]\nonumber \\
 & = & \frac{|z|^{2}}{2}\left|\int dt\int da\, u_{0}u_{0}^{*}\right|^{2}\, ,
\end{eqnarray}
Here we used the result \( \triangle =\frac{n\pi }{2} \) from 
Eq.~(\ref{phasediff}). The momentum dispersion of mirror \( \#2 \),
\( \langle \triangle p_{2}^{2}\rangle  \) will be the same as \( 
\langle \triangle p_{1}^{2}\rangle  \).
The correlation between mirrors is 
\begin{eqnarray}
 &  & \langle \triangle p_{1}\triangle p_{2}\rangle \nonumber \\
 & = & |z|^{2}Re[\int dt\int da\, r\, u_{0}(\sum _{j}r\, u_{j}+\sum 
_{k}t\, u_{k})^{*}\int dt'\int da'\, t^{*}u^{*}_{0}(\sum _{j}t\, 
u_{j}+\sum _{k}r\, u_{k})]\nonumber \\
 & = & |z|^{2}|\int dt\int da\, 
u_{0}u_{0}^{*}|^{2}*\frac{1}{4}Re[(1+e^{i\triangle })(1+e^{i\triangle 
})]\nonumber \\
 & = & |z|^{2}|\int dt\int da\, u_{0}u_{0}^{*}|^{2}*\frac{1}{4}Re[\pm 
2i]\nonumber \\
 & = & 0\, .
\end{eqnarray}
We can see that there is no correlation between arms. The 
fluctuations are
totally independent to each other. The dispersion of the momentum 
difference
becomes 
\begin{equation}
\langle \triangle p^{2}\rangle =2\langle \triangle p_{1}^{2}\rangle 
=|z|^{2}\left(\int dt\int da\, u_{0}u_{0}^{*}\right)^{2}
    =4\langle n\rangle \omega^2 
\, .
\end{equation}
 For \( b \) bounces, the dispersion is
\begin{equation}
\langle \triangle p^{2}\rangle _{b}=2b^{2}\langle \triangle 
p_{1}^{2}\rangle =4b^{2}\langle n\rangle \omega^2 \, .
\end{equation}
This confirms that Eq.~(\ref{eq:v2d_{p}n}) give the correct velocity 
dispersion of each end mirror in the interferometer.

\section{Discussion and Conclusions }
\label{sec:final}

In this paper, we have shown how quantum fluctuations of radiation 
pressure
arise from fluctuations of the stress tensor operator. Our results 
are in agreement
with those obtained previously using a photon number counting 
approach. In our
approach, the radiation pressure fluctuations in a coherent state are 
due entirely
to the cross term in the product of stress tensors. This term is both 
dependent
upon the quantum state, and is singular in the limit of coincident 
points. however,
we found that careful treatment of the integrals over space and time 
leads to a finite result. 

The cross term can be interpreted as representing
the interference between vacuum fluctuations and the real photons
present. Thus radiation pressure fluctuations in the stress tensor
approach are driven by vacuum fluctuations. It is useful to compare
the photon number and stress tensor approaches at this point. Both
approaches yield the same answers for all of the questions which were
posed in this paper. (A possible exception is the Fabry-Perot
cavity discussed in Sect.~\ref{sec:Fabry} using only the  stress tensor 
approach.) However, the conceptual pictures presented by the two approaches
are quite different. In the photon number approach, the pressure fluctuations
on a single mirror are attributed to statistical variations in the numbers 
of photons striking the mirror. However, when one wants to treat the
problem of noise in an intererometer, especially the lack of correlation
between the fluctuations in the two arms, it is necessary to invoke
vacuum fluctuations~\cite{Caves1,Caves2}. In our view, the stress tensor
approach provides a more unified description in which the role of
vacuum fluctuations is clear from the outset. It is also likely to 
generalize more easily to complex situations. For example, all of the
treatments of radiation pressure fluctuations, with which we are aware,
assume that the end mirrors are perfectly reflecting. However, the
stress tensor approach could be easily adapted to account for the
finite reflectivity of this mirror.

Radiation pressure fluctuations will play a role in laser 
interferometer
detectors of gravity waves, especially in the future. At that point, 
it should
become possible to measure these fluctuations experimentally. 
Confirmation of
their existence can be viewed as experimental evidence for the 
reality of the
cross term.

Of special significance is the role of radiation pressure 
fluctuations in understanding
the fundamental physics of stress tensor fluctuations. It seems 
natural that
the same principles which apply to the stress tensor as a source of 
pressure
on a mirror should also apply to the stress tensor as a source of 
gravity.
We have seen that the cross term is essential to understand radiation 
pressure
fluctuations. It then follows that the cross term must be included in 
the treatment
of spacetime metric fluctuations driven by stress tensor fluctuations.

\vspace{0.5cm}

{\bf Acknowledgement:} We would like to thank K. Olum and C. M. Caves
for valuable discussions. This work was supported in part by the National
Science Foundation under Grant PHY-9800965.

\section*{Appendix}

\setcounter{equation}{0} 
\renewcommand{\theequation}{A\arabic{equation}}

In this appendix, we calculate the integral \( J \) defined in Eq.~(\ref{eq:J}).
Define $u=t_{1}-t_{2}$ and $v=t_{1}+t_{2}$. Next use the identities 
\begin{equation}
\cos (\omega t_{1})\cos (\omega t_{2})=\frac{1}{2}[\cos (\omega 
u)+\cos (\omega v)]
\end{equation}
and 
\begin{equation}
\int _{0}^{\tau }\int _{0}^{\tau }dt_{1}dt_{2}=\frac{1}{2}\left( \int 
_{-\tau }^{0}du\int _{-u}^{u+2\tau }dv+\int _{0}^{\tau }du\int 
_{u}^{2\tau -u}dv\right) \, .
\end{equation}
After evaluation of the \( v \)-integrations, we may write 
\begin{equation}
J=\int _{0}^{\tau }du\, \frac{(\tau 
-u)(u^{2}+a)}{(u^{2}-b^{2})^{3}}\, \cos {\omega u}-\frac{1}{2\omega 
}\int _{0}^{\tau }du\, \frac{u^{2}+a}{(u^{2}-b^{2})^{3}}\, [\sin 
{\omega u}+\sin \omega (u-2\tau )]\, .
\end{equation}
The second integral in the above expression approaches a constant as 
\( \tau \rightarrow \infty  \),
whereas the first integral contributes a linearly growing term: 
\begin{equation}
J\sim \tau \int _{0}^{\infty }du\, 
\frac{u^{2}+a}{(u^{2}-b^{2})^{3}}\, \cos {\omega u}=\frac{1}{2}\tau 
\int _{-\infty }^{\infty }du\, \frac{u^{2}+a}{(u^{2}-b^{2})^{3}}\, 
\cos {\omega u}\, .
\end{equation}
This integral contains third-order poles at \( u=\pm b \). It can be 
expressed
as 
\begin{equation}
J=\frac{1}{4}\, \tau (J_{+}+J_{-})\, ,
\end{equation}
where in \( I_{+} \) we assume \( {\textrm{Im}}\, b>0 \) and in \( 
I_{+} \)
we take \( {\textrm{Im}}\, b>0 \). Each of these integrals is in turn 
expressed
as 
\begin{equation}
I_{\pm }=\frac{1}{2}(I_{\pm 1}+I_{\pm 2})\, ,
\end{equation}
where 
\begin{equation}
I_{\pm 1}=\int _{-\infty }^{\infty }du\, 
\frac{u^{2}+a}{(u^{2}-b^{2})^{3}}\, {\textrm{e}}^{i\omega u}
\end{equation}
and 
\begin{equation}
I_{\pm 2}=\int _{-\infty }^{\infty }du\, 
\frac{u^{2}+a}{(u^{2}-b^{2})^{3}}\, {\textrm{e}}^{-i\omega u}\, .
\end{equation}
 Each of these integrals is evaluated by closing the contour 
of integration
in the appropriate half plane, and then evaluating the integral by a 
combination
of integration by parts and Cauchy's theorem. For example, in the 
case of \( I_{+1} \),
we close in the upper half plane and write 
\begin{eqnarray}
I_{+1} & = & \frac{1}{2}\int _{-\infty }^{\infty }du\, \left( 
\frac{d^{2}}{du^{2}}\; \frac{1}{u-b}\right) \, 
\frac{u^{2}+a}{(u+b)^{3}}\, {\textrm{e}}^{i\omega u}=\frac{1}{2}\int 
_{-\infty }^{\infty }du\, \frac{1}{u-b}\, \frac{d^{2}}{du^{2}}\left[ 
\frac{u^{2}+a}{(u+b)^{3}}\, {\textrm{e}}^{i\omega u}\right] \nonumber 
\\
 & = & \pi i\left\{ \frac{d^{2}}{du^{2}}\left[ 
\frac{u^{2}+a}{(u+b)^{3}}\, {\textrm{e}}^{i\omega u}\right] \right\} 
_{u=b}\, .
\end{eqnarray}
Similarly, we find 
\begin{equation}
I_{+2}=-\pi i\left\{ \frac{d^{2}}{du^{2}}\left[ 
\frac{u^{2}+a}{(u-b)^{3}}\, {\textrm{e}}^{-i\omega u}\right] \right\} 
_{u=-b}\, ,
\end{equation}

\begin{equation}
I_{-1}=-\pi i\left\{ \frac{d^{2}}{du^{2}}\left[ 
\frac{u^{2}+a}{(u+b)^{3}}\, {\textrm{e}}^{-i\omega u}\right] \right\} 
_{u=b}\, ,
\end{equation}
and 
\begin{equation}
I_{-2}=\pi i\left\{ \frac{d^{2}}{du^{2}}\left[ 
\frac{u^{2}+a}{(u-b)^{3}}\, {\textrm{e}}^{i\omega u}\right] \right\} 
_{u=-b}\, .
\end{equation}
We may now combine all of these results to obtain the expression for 
\( J \),
Eq. (\ref{eq:J2}).
Note that this calculation involves integrations by parts very 
similar to those used in Ref.~\cite{WF99} and illustrated in
Sect.~\ref{sec: EMTF} of the present paper.


\begin{thebibliography}{}
\bibitem{Caves1}C. M. Caves, Phy. Rev. Lett. \textbf{45}, 75 (1980).
\bibitem{Caves2}C. M. Caves, Phys. Rev. D \textbf{23}, 1693 (1981).
\bibitem{WF99}C.-H. Wu and L.H. Ford, Phys. Rev. D \textbf{60}, 
104013 (1999), gr-qc/9905012.
\bibitem{F82}L.H. Ford, Ann. Phys (NY) \textbf{144}, 238 (1982).
\bibitem{DF88}S. del Campo and L.H. Ford, Phys. Rev. D \textbf{38}, 
3657 (1988).
\bibitem{Kuo}C.-I Kuo and L.H. Ford, Phys. Rev. D \textbf{47}, 4510 
(1993), gr-qc/9304008.
\bibitem{PH97}N.G. Phillips and B.L. Hu, Phys. Rev. D \textbf{55}, 
6123 (1997), gr-qc/9611012.

\bibitem{Davies} K.T.R. Davies and R.W. Davies,
Can. J. Phys. {\bf 67}, 759 (1989); K.T.R. Davies, R.W. Davies,
and G. D. White, J. Math. Phys. {\bf 31}, 1356 (1990).

\bibitem{FJL} D.Z. Freedman, K. Johnson and J.I. Latorre, Nucl. Phys. 
{\bf B371}, 353 (1992).

\bibitem{MF-optics} L. Mandel and E. Wolf, {\it Optical Coherence and 
Quantum Optics} (Cambridge University 
              Press, London, 1995),  Sect.~10.9.5 .
\end{thebibliography}
\end{document}